\documentclass[a4 paper]{article}
\usepackage{epsfig}
\usepackage{multicol}
\usepackage{array}
\usepackage{amsfonts,amsmath,amssymb}

\input{epsf.sty}
\begin{document}
\twocolumn[
\begin{center}
{\Large{\bf Spectral line intensity and polarization in gas-dusty medium}}\\
\bigskip
{N. A. Silant'ev\thanks{E-mail: nsilant@bk.ru}\,\,, G. A. Alekseeva,\, V. V. Novikov}
\medskip\\
\end{center}
\begin{center}
 {Central Astronomical Observatory at Pulkovo of Russian Academy of Sciences,\\ 196140,
Saint-Petersburg, Pulkovskoe shosse 65, Russia\\}
\medskip
\end{center}

\bigskip
\begin{center}
{received..... 2016, \qquad accepted....}
\end{center}]

\begin{abstract}
    We study the standard spectral line radiative transfer equation for media consisting of resonant atoms and non-resonant components
(the dust grains and the atoms without considered spectral transition).
 Our goal is to study  the
intensity and polarization of the resonance radiation emerging from semi-infinite atmosphere. 
 Using the known technique of resolvent matrices, we obtain the exact solution of vectorial radiative transfer equation for
various sources of non-polarized radiation in semi-infinite atmosphere. Homogeneous, linear increasing and exponentially
decreasing sources are considered.  Recall,  that'  with the  homogeneous sources correspond to the isothermal atmosphere and the exponential ones correspond to the incident radiation with
 different angles of inclination. 

\end{abstract}

$^1$ E-mail: nsilant@bk.ru
\section{Introduction}

  The observation of  spectral lines is the most important technique of investigation in astrophysics.  First of all the intensity of spectral line was considered. The angular distribution of scattered radiation was assumed isotropic.  The problems of the scalar spectral line transfer were presented in known monographs  by Ivanov (1973)  and  Mihalas (1982). 

 The problems with  non-isotropic and polarized radiation are more difficult to consider
(see, for example, Stenflo 1976, Rees \& Saliba 1982; Faurobert 1988, Faurobert-Sholl \& Frisch 1989;
Faurobert-Sholl et al. 1997). It should be also noted  the  papers by Ivanov 1996, Ivanov et al. 1997, and  Dementyev 2008, 2010.  Simple quantum mechanical study of the resonance line problems is given in  the standard text-book 
by Landau \& Lifshitz vol. 4,  part 1 (1968) . 

  The polarization of spectral line depends also on the magnetic field.
 Most exhaustive radiative transfer problems in magnetized atmosphere are presented in the books by Fluri (2003) and  Landi degl'Innocenti \& Landolfi (2004).

We present here the  transfer equation for polarized spectral line radiation in non-magnetized media considering three types of particles - atoms with the resonant
level, atoms without considered resonant level and the dust grains.    

 A detailed  quantum-mechanical theory (see Landau \& Lifshitz 1968) gives rise to the following formulas for the  scattering cross-section and the absorption one ( describing the process of collisional deexitation,  or destruction, see Hummer \& Rybicki 1971):
\[
\sigma^{(s)}_{res}(\nu)=\frac{\sigma^{(s)}_{res}}{(\nu-\nu_0)^2+\gamma^2},
\]
\begin{equation}
\sigma^{(a)}_{res}(\nu)\sim \frac{ Const\, \gamma}{(\nu-\nu_0)^2+\gamma^2}.
\label{1}
\end{equation}
\noindent   These cross-sections have the Lorentz shape . Here $\nu_0$ is the central frequency  of resonance line, the value $\gamma $ is the line width. The absorption
coefficient is proportional to small parameter $\gamma $. 
 Note that the  line width includes all types of scattering - both elastic and non-elastic.  Formulas (1) correspond to the rest  frame of the resonance atom.

The allowance for  Doppler shift of frequencies in the thermal motions of atoms gives rise to the final form of cross-sections (see Ivanov 1973):

\[ 
\sigma^{(s)}_{res}(\nu)=\sigma^{(s)}_{res}\,\varphi_V(\nu),
\]
\begin{equation}
\sigma^{(a)}_{res}(\nu)=\sigma^{(a)}_{res}\,\varphi_V(\nu),
\label{2}
\end{equation}
\noindent where $\varphi_V(\nu)$ is known Voigt normalized profile.

Further we will use the total cross-section notation:
\begin{equation}
\sigma^{(t)}\equiv \sigma^{(s)}+\sigma^{(a)}.
\label{3}
\end{equation}

 Very important part of our study is the  choice of redistribution matrix
 $ \hat{R}({\nu,\bf n};{\nu',\bf n'}) $. We  chose the most simple form used in many papers, in  particular,  in  Ivanov et al. 1997.   This is { complete frequency redistribution matrix (see Eqs.(4) and (11)).  The matrix $\hat{R}$
describes the probability to scatter the initial resonance radiation taking the frequency $\nu'$ and direction ${\bf n'}$ into radiation with frequency $\nu$ and direction ${\bf n}$, as a result, of the scattering on a resonant atom.  

  The complete  frequency redistribution matrix has the form:
\begin{equation}
  \hat{R}(\nu,{\bf n};\nu',{\bf n'}) =(1/4\pi)\varphi(\nu)\varphi(\nu')\hat{P}({\bf n},{\bf n'}).
\label{4}
\end{equation}
Here  $ \varphi(\nu)$   is normalized profile of scattering and absorption cross-sections; the matrix $\hat P({\bf n},{\bf n'})$ for a number of cases  is given in the book by Chandrasekhar (1960).

  In radiative transfer equation without introduction of  dimensionless optical depth, the integral term,
descibing the scattering of radiation, contains redistribution matrix with the term $N_{res}\sigma^{(s)}_{res}$. Here $N_{res}$ is the number 
density of resonant atoms, $\sigma^{(s)}_{res}$ is frequency averaged cross-section of the resonance scattering.
Note that the detailed formulas for $\sigma^{(s)}_{res}$ and $\sigma^{(a)}_{res}$ are given in books by Ivanov (1973) and by Mihalas (1982).

 Below we consider the axially symmetric problems, where the Stokes parameter $U=0$. In this case $  \hat{R}(\nu,{\bf n};\nu',{\bf n'}) \to \hat{R}(\nu,\mu;\nu',\mu')$ and the $\hat P({\bf n},{\bf n'})$ transforms to $2\pi\hat P(\mu,\mu')$ form, where $\mu$, $\mu'$ are the cosines of angles between ${\bf n}$ , ${\bf n'}$ and the normal ${\bf N}$ to
the surface of the atmosphere.

\section{Radiative transfer equation for Stokes parameters $I(\tau,\nu,\mu)$ and $Q(\tau,\nu,\mu)$ in the case of spectral line}

We will consider the vectorial radiative transfer equation for the intensity $I$ and
the Stokes parameter $Q$, which describe axially symmetric problems.  
 Usually one  uses the  optical depth:
\begin{equation}
 d\tau=N_{res}(\sigma^{(s)}_{res}+\sigma^{(a)}_{res})dz\equiv N_{res}\sigma^{(t)}_{res}dz,
\label{5}
\end{equation}
\noindent independent of the frequency $\nu$,
where $N_{res}$ is the number density of resonant atoms, $\sigma^{(s)}_{res}$ and $\sigma^{(a)}_{res}$ are the frequency averaged corresponding  cross-sections. Recall, that the cross-sections: 
\begin{equation}
\sigma^{(s,a)}_{res}(\nu)=\sigma^{(s,a)}_{res}\,\varphi(\nu),
\label{6}
\end{equation}
 \noindent where normalized function $\varphi(\nu)$ describes the shape of the scattering and absorption cross-sections.  As the $\varphi(\nu)$- function we take the function, which describes the Doppler shape of spectral line. This shape arises from the Voigt function for very small value of parameter $\gamma$ : 
\[
\varphi_D(\nu)=\frac{1}{\sqrt{\pi}\Delta\nu_D}\exp{\left[-\left ( \frac{\nu-\nu_0}{\Delta\nu_D}\right )^2\right]},
\]
\begin{equation}
\int\limits_{-\infty}^{\infty}d\nu\,\varphi_D(\nu)=1.
\label{7}
\end{equation}
\noindent Here $\Delta\nu_{D}$ is Doppler's width: 

$\Delta\nu_{D}=\nu_0\sqrt{u_{th}^2+u_{turb}^2}/c$, where the thermal velocity along the line of sight is $u_{th}=\sqrt{2kT/M}$, and the analogous turbulent velocity is the mean value of  chaotic motions $u_{turb}^2=\langle {\bf u}^2({\bf r},t)\rangle/3$. Parameter $c$ is the speed of light.
 
  The first  term in radiative transfer equation describes the extinction of the radiation. In our case of the
three - component medium the extiction factor $\chi(\nu)$  has the form:

\[
\chi(\nu)=N_{res}\sigma^{(t)}_{res}\varphi(\nu)+N_{grain}\sigma^{(t)}_{grain}+N_{atom}\sigma^{(t)}_{atom}\equiv
\]
\begin{equation}
 N_{res}\sigma^{(t)}_{res}\alpha(\nu),
\label{8}
\end{equation}
\noindent where dimensionless extinction factor $\alpha(\nu)$ is:

\[
\alpha(\nu)=\varphi(\nu)+\beta, 
\]
\begin{equation}
\beta=\frac{N_{grain}\sigma^{(t)}_{grain}+N_{atom}\sigma^{(t)}_{atom}}{N_{res}\sigma^{(t)}_{res}}\equiv \beta_{grain}+\beta_{atom}.
\label{9}
\end{equation}
\noindent Parameter $\beta$ descibes the extinction by dust grains and non-resonant atoms, i.e. by particles which do not have the considered resonant level. $N_{grain}$ and  $N_{atom}$ are corresponding number densities.

According to usual procedure of derivation of radiative transfer equation with the optical depth $d\tau=N_{res}\sigma^{(t)}_{res}dz$, we obtain the following equation: 
  
\[
\mu\frac{d{\bf I}(\tau,\nu,\mu)}{d\tau}=\alpha(\nu){\bf I}(\tau,\nu,\mu)-
\]
\[
(1-\epsilon)\int\limits_{-1}^{1}
d\mu'\int\limits_{-\infty}^{\infty}d\nu'\,\hat{R}(\nu,\mu;\nu',\mu'){\bf I}(\tau,\nu',\mu')-
\]
\begin{equation}
\varphi(\nu)s(\tau,\nu)\left (\begin {array}{c} 1 \\ 0 \end{array} \right ).
\label{10}
\end{equation}
\noindent Here we introduced the (column) vector {\bf I} with the components ($I,Q$), where $I(\tau,\nu, \mu)$ and $Q(\tau,\nu, \mu)$ [erg/cm$^{2}$ Hz s sr] are the intensity and the Q- Stokes parameter, respectively. The value
$\nu$ is the frequency of light, $\mu=\cos\vartheta$. The angle $\vartheta$ is the angle between the  line of sight ${\bf n}$ and the normal to the plane-parallel atmosphere ${\bf N}$.  The parameter $(1-\epsilon)=\sigma^{(s)}_{res}/\sigma^{(t)}_{res}$   is the  probability of scattering on the resonant atom. Parameter $\epsilon=\sigma^{(a)}_{res}/\sigma^{(t)}_{res}$ is the destruction probability  (see Frisch \& Frisch  1977).  The source term $\varphi(\nu)s(\tau,\nu)$
is the scattered non-polarized isotropic  Plank's radiation. 
 
 The cross-section of the resonance line scattering in
optical range is of the order $\sigma_{res}\simeq 10^{-16}$ cm$^2$ (see, for example, Gasiorovich 1996). The cross-section of the scattering for  non-resonant radiation is
much lesser $\sigma_{atom}\simeq 10^{-28}$ cm$^2$. Remind, that the Thomson cross-section of scattering on free electrons is $\sigma_T\simeq 6.4\cdot10^{-25}$ cm$^2$, larger than atom's cross-section. The cross-section $\sigma_{grain}\sim r^2>>\sigma_{atom}$, where $r$ is the radius of a grain.   It appears that $\sigma^{(t)}_{grain}>>\sigma^{(t)}_{atom}$.

  It is also  seen from Eq.(9) that the extinction factor $\beta=\beta_{grain}+\beta_{atom}$ is the sum of the part, depending on the dust grains, and part, depending on non-resonant atoms.
The case $N_{atom}>>N_{grain}$ may correspond to $\beta_{atom}\sim 
\beta_{grain}$. We will investigate the dependence of the radiation intensity and polarization on the parameter $\beta$ which consists of
the sum of these parameters. Note that  only the particular models can present the values $\beta_{grain}$ and 
$\beta_{atom}$ separately.

The matrix $\hat R(\nu,\mu;\nu',\mu')$ in general has very complex form
 
(see McKenna 1985: Landi degl'Innocenti \& Landolfi 2004). 

This is the reason why one uses the model of complete (full) redistributed matrix:

\begin{equation}
\hat{R}(\nu,\mu;\nu',\mu')=\frac{1}{2}\varphi(\nu)\varphi(\nu')\hat{P}(\mu,\mu'),
\label{11}
\end{equation}
\noindent  It appears that formula (11) corresponds to the case when during the  lifetime of atomic level  many impacts hold.

 The matrix $\hat{P}(\mu,\mu')$ has the form:

\begin{equation}
\hat{P}(\mu,\mu')=\hat A(\mu^2)\hat A^T(\mu'^2).
\label{12}
\end{equation}

 The matrix $\hat A(\mu^2)$ is (see Ivanov et al. 1997):
\begin{equation}
\hat A(\mu^2)= \left (\begin{array}{rr}1 ,\,\, \sqrt{\frac{W}{8}}(1-3\mu^2) \\ 0 , \,\,3\sqrt{\frac{W}{8}}(1-\mu^2) \end{array}\right).
\label{13}
\end{equation}
\noindent  Here parameter $W$ depends on quantum numbers of transition atomic levels. For simplest case of dipole transition $W=1$.  
Our calculations in Ch. 5 corresponds to this case.
 Note that the superscript T will be used for matrix transpose.
Note also that there exists the equality:
\begin{equation}
\left (\begin {array}{c} 1 \\ 0 \end{array} \right )\equiv  \hat{A}(\mu^2)\left (\begin {array}{c} 1 \\ 0 \end{array} \right ).
\label{14}
\end{equation}
 Using the dimensionless frequencies $x=(\nu-\nu_0)/\Delta\nu_D$ and identity (14) , the transfer equation (10) can be written in the form:

\begin{equation}
\mu\frac{d{\bf I}(\tau,x,\mu)}{d\tau}=\alpha(x){\bf I}(\tau,x,\mu)-\varphi(x)\hat A(\mu^2){\bf S}(\tau),
\label{15}
\end{equation}
\noindent  where the vector ${\bf S}(\tau)$ is:

\[
{\bf S}(\tau)=s(\tau)\left (\begin{array}{c}1\\0 \end{array}\right) +
\]
\begin{equation}
  \frac{1-\epsilon}{2}\int\limits_{-1}^1\,d\mu\int\limits_{-\infty}^{\infty}dx\varphi(x)\hat A^T(\mu^2){\bf I}(\tau,x,\mu).
\label{16}
\end{equation}

 It is easy check that for the case $\beta=0$, $\epsilon=0$ and $s(\tau)=0$ there exists the conservation law for the total flux of radiation:
\begin{equation}
\frac{dF(\tau)}{d\tau}=0,\,\, F(\tau)=\int_{-1}^{1}d\mu\int_{-\infty}^{\infty}dx\,I(\tau,x,\mu).
\label{17}
\end{equation}

Using known formal solution of Eq.(15) (see Chandraserhar 1960, Silant'ev et al. 2015), we derive the integral equation for ${\bf S}(\tau)$:

\begin{equation}
{\bf S}(\tau)={\bf g}(\tau)+\int\limits_0^{\infty} d\tau'\hat L(|\tau-\tau'|){\bf S}(\tau').
\label{18}
\end{equation}
\noindent  The free term ${\bf g}(\tau)$ has the form:
\begin{equation}
{\bf g}(\tau)= s(\tau)\left (\begin{array}{c}1\\0 \end{array}\right).
\label{19}
\end{equation}
 The matrix kernel of  integral equation (18) is the following:

\[
\hat L(|\tau-\tau'|)=\int\limits_0^1\frac{d\mu}{\mu}\int\limits_{-\infty}^{\infty}dx\varphi^2(x)\times
\]
\begin{equation}
\exp{\left(-\frac{\alpha(x)|\tau-\tau'|}{\mu}\right)}\,\hat \Psi(\mu^2),
\label{20}
\end{equation}
\noindent where
\begin{equation}
 \hat \Psi(\mu^2)=\frac{1-\epsilon}{2}\hat A^T(\mu^2)\hat A(\mu^2).
\label{21}
\end{equation}
\noindent Note that the matrix $\hat \Psi$ is symmetric: $\hat \Psi^T=\hat \Psi$. This property gives rise to symmetry of kernel $\hat L^T=\hat L$. The explicit form of matrix $\hat \Psi(\mu^2)$ is the following:
\[
\hat\Psi(\mu^2)=
\]
\begin{equation}
\frac{1-\epsilon}{2} \left (\begin{array}{rr}1,\qquad \,\,\,  \sqrt{\frac{W}{8}}(1-3\mu^2) \\ \sqrt{\frac {W}{8}}(1-3\mu^2), \,\, \frac{W}{4}(5-12\mu^2+9\mu^4) \end{array}\right).
\label{22}
\end{equation}

Below we follow to general theory of resolvent matrices, which is given in Silant'ev et al. (2015).

\section{Solution of integral equation for ${\bf S}(\tau)$ using resolvent matrix}

 According to the standard theory of integral equations (see, for example, Smirnov 1964), the solution of  Eq. (18) can be presented in the form:

\begin{equation}
{\bf  S}(\tau)={\bf g}(\tau)+\int\limits_0^{\infty}d\tau'\hat {R}(\tau,\tau'){\bf  g}(\tau'),
\label{23}
\end{equation}
\noindent where the resolvent matrix $\hat{R}(\tau,\tau')$ obeys the integral equation
\begin{equation}
\hat{R}(\tau,\tau')=\hat{L}(|\tau-\tau'|)+\int\limits_0^{\infty}d\tau''\hat L(|\tau-\tau''|)\hat{R}(\tau'',\tau').
\label{24}
\end{equation}
\noindent It has the property $\hat{R}^T(\tau,\tau')=\hat{R}(\tau',\tau)$.  We see that 
the equation for $\hat{R}(\tau,0)$ follows from Eq.(24):

\begin{equation}
\hat{R}(\tau,0)=\hat{L}(\tau)+\int\limits_0^{\infty}d\tau'\hat L(|\tau-\tau'|)\hat{R}(\tau', 0).
\label{25}
\end{equation}

The general theory (see Sobolev (1969) and  Silant'ev et al. 2015) demonstrates that the resolvent $\hat{R}(\tau,\tau')$ can be calculated, if we know the martices $\hat{R}(\tau,0)$ and $\hat{R}(0,\tau')$. This is seen directly from the expression for double Laplace transform of $\hat{R}(\tau,\tau')$ with the parameters $a$ and $b$:

\begin{equation}
\tilde{\tilde{\hat{R}}}(a,b)=\frac{1}{a+b}[\,\tilde{\hat{R}}(a,0)+\tilde{\hat{R}}(0,b)+\tilde{\hat{R}}(a,0)\tilde{\hat{R}}(0,b)].
\label{26}
\end{equation}

Taking the  Laplace transform of $\hat{R}(\tau,0)$  and using the relation (26), we can derive non-linear equation for $H$ -matrix:

\begin{equation}
\hat{H}(z)=\hat{E} +\tilde{\hat{R}}\left(\frac{1}{z},0\right)\,\,\,,
\label{27}
\end{equation}
\noindent where $ \tilde{\hat{R}}(1/z,0)$ is the Laplace transform of  $\hat {R}(\tau,0)$ with parameter $1/z$.
 $\hat{E}$  is the unit matrix. This equation has the form:

\[
\hat{H}(z)=\hat{E} +
\hat{H}(z)\int_{-\infty}^{\infty}dx'\varphi^2(x')\int_0^1\frac{d\mu'}{\mu'}\times
\]
\begin{equation}
\frac{1}{1/z+\alpha(x')/\mu'}\hat{H}^T\left(\frac{\mu'}{\alpha(x')}\right)\hat{\Psi}(\mu'^2),
\label{28}
\end{equation}

The  $\hat{H}(z)$-matrix can be calculated, if we know $\hat{H}\left(\frac{\mu}{\alpha(x)}\right)$, which obeys the following non-linear equation:

\[
\hat{H}\left(\frac{\mu}{\alpha(x)}\right)=\hat{E} +\hat{H}\left(\frac{\mu}{\alpha(x)}\right)\int_{-\infty}^{\infty}dx'\int_0^1\frac{d\mu'}{\mu'}\times
\]
\begin{equation}
\frac{\varphi^2(x')}{\alpha(x)/\mu+\alpha(x')/\mu'}\hat{H}^T\left(\frac{\mu'}{\alpha(x')}\right)\hat\Psi(\mu'^2).
\label{29}
\end{equation}

The effective methods of numerical calculation of $\hat{H}$ -matrix  are presented in papers Krease \& Siewert (1971) , Rooij et al.(1989) and Dementyev (2008).

\section{Formulas for emerging radiation}

According to Eq.(15), the vector ${\bf I}(0,x,\mu)$, describing the outgoing radiation, has the form:
\[
{\bf I}(0,x,\mu)\equiv{\bf I}(x,\mu)=\varphi(x)\hat {A}(\mu^2)\times
\]
\begin{equation}
\int\limits_0^{\infty}\frac{d\tau }{\mu}\exp{\left (-\frac{\alpha(x)\tau}{\mu}\right )}{\bf S}(\tau),
\label{30}
\end{equation}
\noindent  i.e. this expression is  proportional to the Laplace transform of ${\bf S}(\tau)$ over variable $\tau $.
   The presence of function $\varphi(x)\sim \exp{(-x^2)} $ in Eq.(30) guarantees that ${\bf I}(x,\mu)$ tends to zero
for $x\sim (3\div4)$.
 The vector ${\bf  S}(\tau)$  is presented in Eq.(23). The Laplace transform of this vector can be written as:

\[
\tilde{{\bf  S}}\left (\frac{\alpha(x)}{\mu}\right )=
\]
\begin{equation}
 \int\limits_0^{\infty}d\,\tau\left [ \hat  E \exp{\left (-\frac{\alpha(x)\tau}{\mu}\right )}+\tilde{\hat  R}\left (\frac{\alpha(x)}{\mu},\tau\right)\right ]{\bf  g}(\tau).
\label{31}
\end{equation}
\noindent  Thus, expression (30) acquires the form:
\[
{\bf  I}(x,\mu)=\varphi(x)\hat A(\mu^2)\frac{1}{\mu}
\left [\tilde {\bf  g}\left (\frac{\alpha(x)}{\mu}\right ) +\right.
\]
\begin{equation}
\left. \int\limits_0^{\infty} d\,\tau
\tilde{\hat {R}}\left (\frac{\alpha(x)}{\mu},\tau\right ){\bf  g}(\tau)\right ].
\label{32}
\end{equation}

  Eq.(32) is the general expression for outgoing radiation, where the term $s(\tau)$ characterizes the distribution of sources of non-polarized radiation in an atmosphere. Explicit form of ${\bf g}(\tau)$ can be obtained from the model of an atmosphere (see Eq.(19)).  It appears, the distribution of $s(\tau)$ (i.e. the model of an atmosphere) can be approximated in the form:
\begin{equation}
s(\tau)\simeq \sum_{n}s_h^n\exp{(-h_n\tau)}+s_0+s_1\tau+s_2\tau^2+...
\label{33}
\end{equation}

For source function $s_h(\tau)=s_h\exp{(-h\tau)}$ the expression for ${\bf I}(x,\mu)$ acquires comparatively simple form. In this case the expression for ${\bf I}(x,\mu)$ depends on $\hat {H}(\alpha(x)/\mu)$ and $\hat H^T(\alpha(x)/\mu)$, i.e. it does not depend on the total matrix $\hat {R}(\tau,\tau')$.

The sources of types $s_n(\tau )=s_n\tau^n$ are related with exponential source by the simple formula:
\begin{equation}
s_n(\tau)=(-1)^n s_n \frac{d^n}{dh^n}\exp{(-h\tau)}|_{h=0} .
\label{34}
\end{equation}
It means that the sources of type (33)
can be considered  on the base of the exponential source.

Let us consider this case in detail. Taking $s_h(\tau)=s_h\exp{(-h\tau)}$  in Eq.(31), we obtain:

\[
{\bf I}_h(x,\mu)=s_h\varphi(x)\hat{A}(\mu^2)\times
\]
\begin{equation}
\hat{H}\left(\frac{\mu}{\alpha(x)}\right)\frac{\hat{H}^T\left(\frac{1}{h}\right)}{\alpha(x)+\mu h}\left (\begin {array}{c} 1 \\ 0 \end{array} \right ).
\label{35}
\end{equation}

 Homogeneous source $s_0$ corresponds to $h=0$.  Physically this case corresponds to homogeneous isothermal atmosphere.  In this case we obtain from Eq. (35):

\[
{\bf I}_0(x,\mu)=s_0\frac{\varphi(x)}{\alpha(x)}\hat{A}(\mu^2)\times
\]
\begin{equation}
\hat{H}\left(\frac{\mu}{\alpha(x)}\right) \hat{H}^T(\infty)
\left (\begin {array}{c} 1 \\ 0 \end{array} \right ).
\label{36}
\end{equation}

 The  emerging radiation for the source $s_1\tau$ can be calculated following the formulas in Silant'ev et al. (2015).

\section{The results of calculations}

It is easy to check (see, for example, Ivanov 1973)  that Eq.(29) for matrix $\hat {H}(\mu/\alpha(x))$ can be presented as the equation for
variable $z=\mu/\alpha(x)$. The emerging radiation (see Eq. (35)) depends on the product $\hat{A}(\mu^2)\hat{H}(\mu/\alpha(x))\equiv \hat{H}_0(x,\mu)$ . We prefer to derive the equation for $\hat{H}_0(x,\mu)$. In this case we are not deal with
the complex problem of approximation of $\hat{H}(\mu/\alpha(x))$ from the value $\hat{H}(z)$. From Eqs.(21) and (29) we derive the following equation for $\hat{H}_0(x.\mu)$:

\[
\hat{H}_0(x,\mu)=\hat{A}(\mu^2) +(1-\epsilon)\hat{H}_0(x,\mu)\mu\int_{0}^{\infty}dx'\int_0^1d\mu'\times
\]
\begin{equation}
\frac{\varphi^2(x')}{\alpha(x)\mu'+\alpha(x')\mu}\hat{H}_0^T(x',\mu')\hat{A}(\mu'^2).
\label{37}
\end{equation}
\noindent  Here we take into account that $\hat{H}_0(x^2,\mu)$ depends on $x^2$. For this reason we take  the $x'$ - integration in the interval $ (0,+\infty)$ . It is clearly that  Eq.(37)  depends on  parameters $\beta$ and $(1-\epsilon)$.  Recall, that ($\alpha(x)=\varphi(x^2)+\beta$). 

 The solution of Eq.(37) can be carry out  by iteration method, analogous to known Chandrasekhar's (1960) method (see also Dementyev (2008)).   For values of parameter $\beta$ we chose
 $\beta=0.001, 0.01, 0.05.$ 
$ 0.1, 0.2, 0.3... 0.9$.

 Note that our calculations correspond to the case $\epsilon=0$. Recall, that in astrophysical conditions parameter $\epsilon$  is
small $\epsilon\simeq$ 10$^{-4}$ (see Ivanov 1973, Frisch \& Frisch 1977).

\subsection{The shapes of resonance line}

The shapes of  resonance line
 
  $J(x)=F_I(x)/F_I(0)$
  are given in Fig.1 (for homogeneous $s_0$ and linearly increasing $s_1\tau$ sources), and in Fig.2 ( for exponentially decreasing  sources $ s(h=1)\exp{(-\tau)} $ and  $ s(h=2)\exp{(-2\tau)})$. In calculations we
assume that all coefficients $s_0, s_1, s(h=1)$ and $s(h=2)$ are equal to unity.
Here the function $F_I(x)$ is the flux of  intensity $I(x,\mu)$ :

\begin{equation}
F_I(x)=\int_0^1 d\mu\, \mu\, I(x,\mu).
\label{38}
\end{equation}

 The value $F_I(x)$  presents the intensity flux  as the function of the dimensionless frequency $x$.
 Recall, that the flux $F_I(x)$   one observes  from the distant spherical  stars (really one observes the value

 $2\pi R^2_{\odot}F_I/R^2$, where $R$ is the distance to a  star and $R_{\odot}$ is the radius of a star).

 The linear polarization of radiation from the spherical star is equal to zero due to axial symmetry of the problem. The Figures 1 and 2 -  demonstrate the dependence of spectral line shape on the parameter $\beta$.
\begin{figure*}[t!]
\includegraphics[width=16cm]{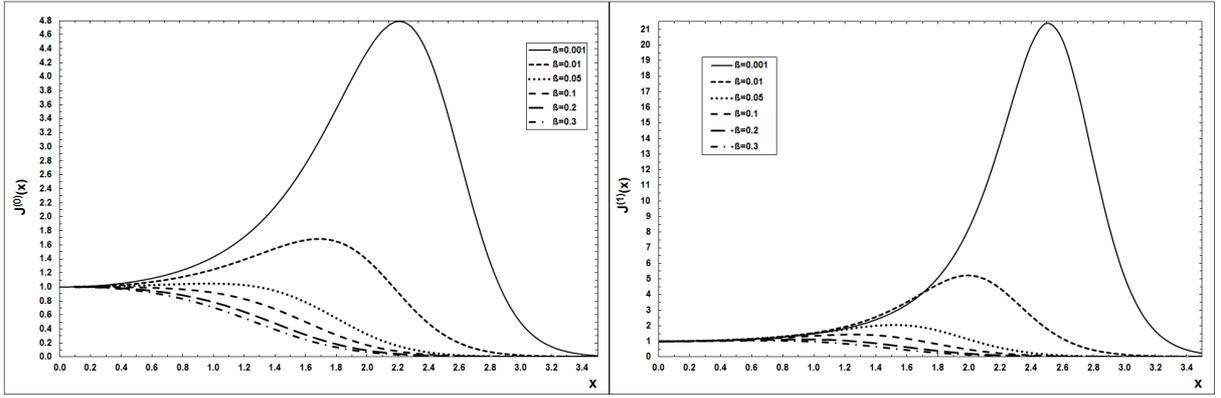}
\caption{The shapes of spectral line 
$J(x)=F_I(x)/F_I(0)$
 for homogeneous source (left curves) and for linearly increasing  source (right curves) for different values of parameter $\beta$.}
 \label{a}
\end{figure*}

\begin{figure*}[t!]
\includegraphics[width=16cm]{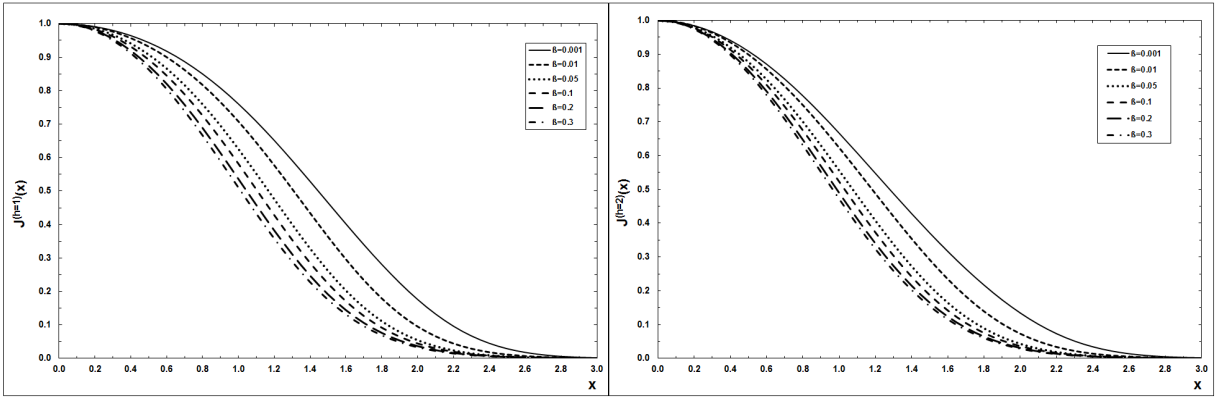}
\caption{The shape of spectral line $J(x)=F_I(x)/F_I(0)$ for source $\exp{(-\tau)}$ (left curves) and for  source  $\exp{(-2\tau)}$ (right curves) for different values of parameter $\beta$.}
 \label{a}
\end{figure*}

 The homogeneous  sources with small parameters $\beta=0.001$ and $0.01$  give rise to the  minimum of outgoing radiation at $x=0$.
In these cases  the resonance line near $x=0$  looks like an absorption line. For $\beta=0.001$ the maximum value of $J(x)$  is  $\simeq 4.8$ and   corresponds to $x\simeq 2.1$. For $\beta=0.01$ the maximum of the line is considerably less $\simeq 1.7$ and corresponds to $x\simeq 1.6$.  For $\beta=0.05 ,0.1, 0.2$ and $0.3$ the shape of the  line looks like usual line profile $\sim \exp{(-x^2)}$ with the maximum at $x=0$. 

 The linearly growing sources  are result in qualitatively  the same behavior, but the maxima of the lines are more profound ($J(x) \simeq 20.4$ at $x\simeq 2.5$ for $\beta=0.001$) and ($\simeq 5.2$ at $x\simeq 2.0$ for $\beta=0.01$ .) 

 The presence of $\varphi(x)$ in Eq.(30) means that the intensity of the resonance line tends to zero for $x> 3\div4$. Why does the peak  of  the resonance  line  arise for small values of absorption parameter $\beta$ ?
 Let us  bear in mind that the radiation density $S_I(\tau)$ increases due to diffusion of
radiation  with the growing $\tau$. For small $\beta$  the emerging radiation goes from the regions far from the surface, where radiation
density is much larger than that near the surface.

The exponential sources $\sim \exp{(-n\tau)}$ are close to the boundary of a medium and the diffusion  of radiation does not give rise to
large radiation density far from the boundary. As a result,  the peaks do not arise.  Fig.2 shows this.

 The absolute values of intensity flux can be obtained from the function $J(x)$  if we know the flux $F_I(0)$ and the constants $ s_n=s_0, s_1 ,s(h=1), s(h=2)...$:

\begin{equation}
F_I(x)=s_nF_I(0)J(x).
\label{39}
\end{equation}
\noindent  The values $F_I(0)$ we present in Tables 1 and 2 for a number of parameters $\beta$ . The value
$F_I(0)$  presents the flux in the centre of a spectral line.
In these Tables  we also present the integrals:
\begin{equation}
\Phi_{int}=\int_0^4dxF_I(x).
\label{40}
\end{equation}
\noindent It is seen, that the values $F_I(0)$ and $\Phi_{int}$ decrease with the increasing of parameter $\beta$.

\subsection{Integral angular distribution and polarization from optically thick accretion discs}

\begin{table}[h!]
\caption{ The values of $F_I(0)$ and $\Phi_{int}$  for homogeneous sources  and sources $\sim \tau$ .}
\small
\begin{tabular}{p{1.1cm}|p{1.1cm}p{1.1cm}|p{1.1cm}p{1.1cm}}
\hline
$\beta$       & $F^{(0)}_{I}(0)$     & $\Phi^{(0)}_{int}$   & $F^{(1)}_{I}(0)$  & $\Phi^{(1)}_{int}$  \\
\hline
  0.001   & 12.283     & 82.49       & 839.1   & 16240.      \\
  0.01     & 4.107       & 12.05       & 44.86   & 284.35    \\
  0.05     & 1.803       & 3.227       & 6.377   & 19.604    \\
  0.1       & 1.218       & 1.813       & 2.773   & 6.368     \\
  0.2       & 0.793       & 1.003       & 1.177   & 2.059     \\
  0.3       & 0.603       & 0.703       & 0.695   & 1.052     \\
  0.4       & 0.491       & 0.543       & 0.470   & 0.648    \\
  0.5       & 0.416       & 0.444       & 0.343   & 0.442     \\
  0.6       & 0.361       & 0.376       & 0.263   & 0.322      \\
  0.7       & 0.320       & 0.326       & 0.208   & 0.246     \\
  0.8       & 0.287       & 0.288       & 0.170   & 0.194     \\
  0.9       & 0.261       & 0.258       & 0.141   & 0.157     \\
\hline
\end{tabular}
\end{table}

\begin{table}[h!]
\caption{ The values of $ F_I(0)$ and  $\Phi_{int}$  for sources $\sim \exp{(-\tau)}$ and $\sim \exp{(-2\tau)}$.}
\small
\begin{tabular}{p{0.8cm}|p{1.2cm}p{1.2cm}|p{1.2cm}p{1.2cm}}
\hline
$\beta$    & $ F^{(h=1)}_I(0)$   & $\Phi^{(h=1)}_{int}$   & $ F^{(h=2)}_I(0)$  &  $\Phi^{(h=2)}_{int}$  \\
\hline
 0.001     & 0.643        & 0.901    & 0.361    & 0.459    \\
  0.01      & 0.585        & 0.744    & 0.336    & 0.393   \\
  0.05      & 0.480        & 0.552    & 0.288    & 0.310    \\
  0.1        & 0.414        & 0.453    & 0.257    & 0.266   \\
  0.2        & 0.338        & 0.353    & 0.220    & 0.220   \\
  0.3        & 0.292        & 0.296    & 0.197    & 0.192     \\
  0.4        & 0.259        & 0.258    & 0.179    & 0.173     \\
  0.5        & 0.234        & 0.230    & 0.166    & 0.158    \\
  0.6        & 0.214        & 0.208    & 0.155    & 0.146     \\
  0.7        & 0.198        & 0.190    & 0.145    & 0.136   \\
  0.8        & 0.184        & 0.175    & 0.137    & 0.128     \\
  0.9        & 0.172        & 0.163    & 0.130    & 0.121     \\
\hline
\end{tabular}
\end{table}

 If we observe by a telescope the intensity and polarization of radiation from the distant optically thick  accretion disc, we observe the radiation fluxes:

\[
F_I^{(Tel)}(x,\mu)=\frac{S}{R^2}\,\mu\, I(x,\mu)\,
\]
\begin{equation}
F^{(Tel)}_Q(x,\mu)=\frac{S}{R^2}\,\mu\, Q(x,\mu)\,.
\label{41}
\end{equation}
\noindent Here $R$ is the distance to a disc, $\vartheta$ is the angle between the line of sight ${\bf n}$ and the normal to a disc ${\bf N}$; $\mu=\cos\vartheta$, S - is the observed surface of homogeneous disc.

 In Figs. $3\div6$ we give the angular distributions $J_{int}(\mu)$ and degrees of polarization $p_{int}(\mu)\%$ for all considered types of sources,  when the  telescope observes the radiation  in the $x$-interval (0,4), as a whole:

\[
F_I^{(Tel)}(\mu)=\frac{S}{R^2}\,\mu\,I_{int}(\mu)\,,
\]
\begin{equation}
F_Q^{(Tel)}(\mu)=\frac{S}{R^2}\,\mu\,Q_{int}(\mu)\,,
\label{42}
\end{equation}
\noindent where $I_{int}(\mu)$ and $Q_{int}(\mu)$ are:
\[
I_{int}(\mu)=\int_0^4dx\,I(x,\mu),
\]
\begin{equation}
Q_{int}(\mu)=\int_0^4dx\,Q(x,\mu).
\label{43}
\end{equation}
\noindent The angular distribution of integral radiation has the following definition:

\begin{equation}
J_{int}(\mu)=\frac{I_{int}(\mu)}{I_{int}(0)}.
\label{44}
\end{equation}

The degree of polarization $p_{int}(\mu)$ is determined  by the ratio:
\begin{equation}
p_{int}(\mu)=\frac{Q_{int}(\mu)}{I_{int}(\mu)}.
\label{45}
\end{equation}

  Recall, that negative polarization denotes that the electric field of electromagnetic wave oscillates perpendicular to the plane $({\bf nN})$, i.e. parallel to the disc's plane. Such oscillations take place at multiple scattering of radiation on free electrons (the Milne problem, see Chandrasekhar 1960).

The  integral values (44) and (45)  we observe by  the telescope which does not resolve
particular frequencies inside the spectral line. Recall, that frequencies $x>4$  are blanketed by continuum radiation and by the wings of neighboring spectral lines. The values $J^{(n)}_{int}(\mu)$ and $p^{(n)}_{int}(\mu)$ are given in Figures $3\div6$.

In  Table 3 we give  the $I_{int}(0)$ -values for a number of  considered types of sources. It is seen that all values decrease with the increasing of the parameter $\beta$. 
\begin{figure*}[t!]
\includegraphics[width=16cm]{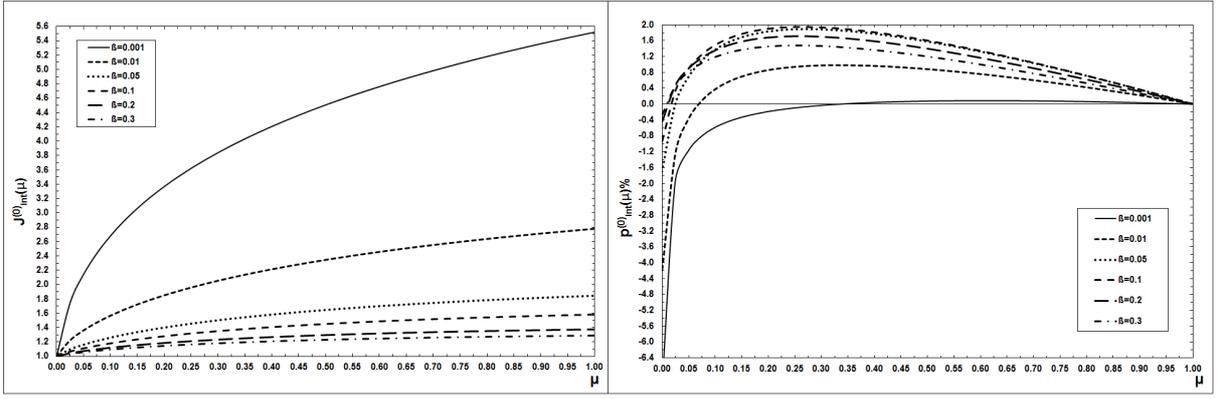}
\caption{Angular distribution $J_{int}(\mu)$ and degree of polarization $p_{int}(\mu)\%$ for homogeneous source  for different values of parameter $\beta$. Negative $p_{int}(\mu)$ corresponds to wave electric field oscillations  parallel to the accretion disc. }
 \label{a}
\end{figure*}

\begin{figure*}[t!]
\includegraphics[width=16cm]{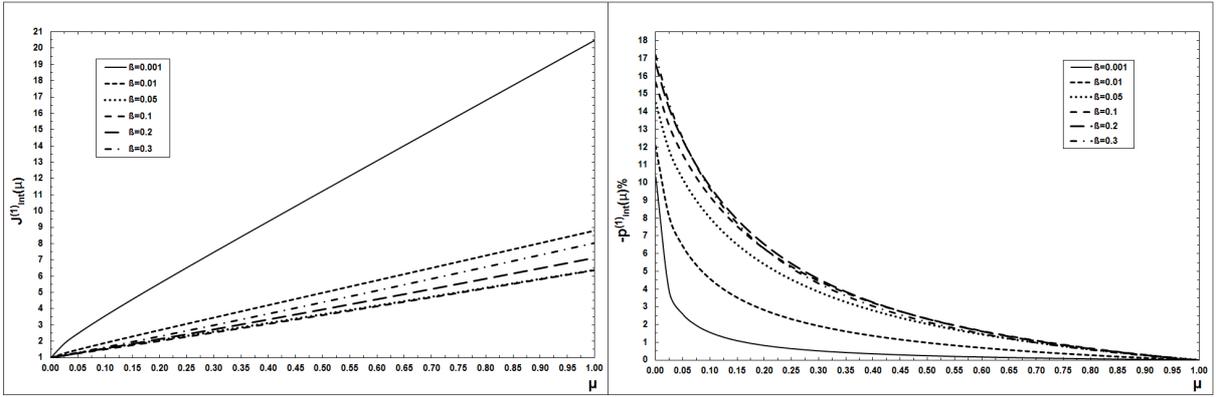}
\caption{Angular distribution $J_{int}(\mu)$ and degree of polarization  (we present the value - $p_{int}(\mu)\%$) for linearly increased source  for different values of parameter $\beta$. Here wave electric field oscillations allways parallel to the accretion disc. }
 \label{a}
\end{figure*}

For homogeneous sources the angular distribution $J_{int}(\mu)$  increases monotomically with $\mu \to 1$. For $\beta=0.001$  we
have $J_{int}(1)\simeq 5.52$. For $\beta=0.01$ and $0.1$ the increase is less -$J_{int}(1)\simeq 2.78$ and $1.58$, respectively.
For larger values of $\beta$ the angular distribution tends to be isotropic. Such behavior of angular distribution of a spectral line is
opposite to angular distribution of continuum radiation for the Milne problem, where for  $\epsilon=0$ we have
$J(1)\simeq 3.06$ and for $\epsilon=0.1$ one has $J(1)\simeq 4.39 $ (see Chandrasekhar 1960; Silant'ev 1980). It appears, this behavior
is due to the presence of radiation peaks  for small values of $\beta$ (the less is parameter $\beta$, the greater is the intensity of
outgoing radiation). 

The total picture of angular distribution depends on two reasons - the presence of peaks and the usual dependence of emerging radiation $\sim \exp{(-\tau/\mu)}$. The latter gives rise to greater intensity at $\mu=1$ compared to $\mu=0$. Recall, that
the  outgoing radiation of the resonance line arises at different $\tau$, depending on the frequency $x$.

The polarization of outgoing radiation mostly arises  by  last scattering before escape from an atmosphere. If the incident radiation is parallel to ${\bf N}$, then degree of polarization $p\sim (1-\mu^2)$ and the wave electric field
oscillates perpendicular to scattering plane $({\bf nN})$. If the incident radiation is perpendicular to ${\bf  N}$, then the oscillations of  integral radiation corresponds  to oscillations in the plane $({\bf nN})$. Of course, the total polarization also depends on angle
$\vartheta (\cos\vartheta=\mu)$. In general, when the  intensity of incident radiation along ${\bf N}$ is much greater than that  in perpendicular direction,  the total polarization of outgoing radiation has oscillations perpendicular to the scattering plane $({\bf nN})$,
i.e. are parallel to the accretion disc. More detail consideration is presented in  Dolginov et al. (1995).  Just such behavior we observe in our case. For resonance line at $\beta=0.001$ the radiation
is directed mostly along ${\bf N}$ and, as a result, we have large degree of negative polarization $p_{int}(0)=-7.17\%$. For $\beta =0.1, 0.2$ and $0.3$, when the incident radiation is mostly perpendicular to the  plane ${\bf nN}$ , the wave electric field oscillates in the plane
$({\bf nN})$. However, this polarization is not high and acquires the maximum value $\simeq 1.9\%$ for the case $\beta=0.05$.

For linearly increasing source the angular distribution $J_{int}(\mu)$ has more complicated behavior than that in the case of homogeneous source. Every curve   $J_{int}(\mu)$ monotonically  increases with $\mu\to1$. However, the $\beta$ - dependence differs from that for  homogeneous $J_{int}(\mu)$. Firstly the value  $J_{int}(1)$  decreases monotonically with the growth of $\beta$ from the value $20.50$ at $\beta=0.001$ to $6.38$ at $\beta=0.1$. Then value  $J_{int}(1)$ begins increase up to value $8.05$ at $\beta=0.3$. It appears, both mechanisms, mentioned above, work more intensive because the density of radiation $S_I(\tau)$ increases more rapidly with the optical depth $\tau$ than in the case of homogeneous sources.
  We see that all angular distributions have very elongated form. As a result, the polarization $p_{int}(\mu)$ is negative for every
value $\mu$, i.e. the wave electric field oscillations are perpendicular to the plane $({\bf nN})$. The polarization $p_{int}(\mu)$
acquires maximum value at $\mu=0$. These values  are within the limits (- $10.3\% \div-17.2\%$.)
\begin{figure*}[t!]
\includegraphics[width=16cm]{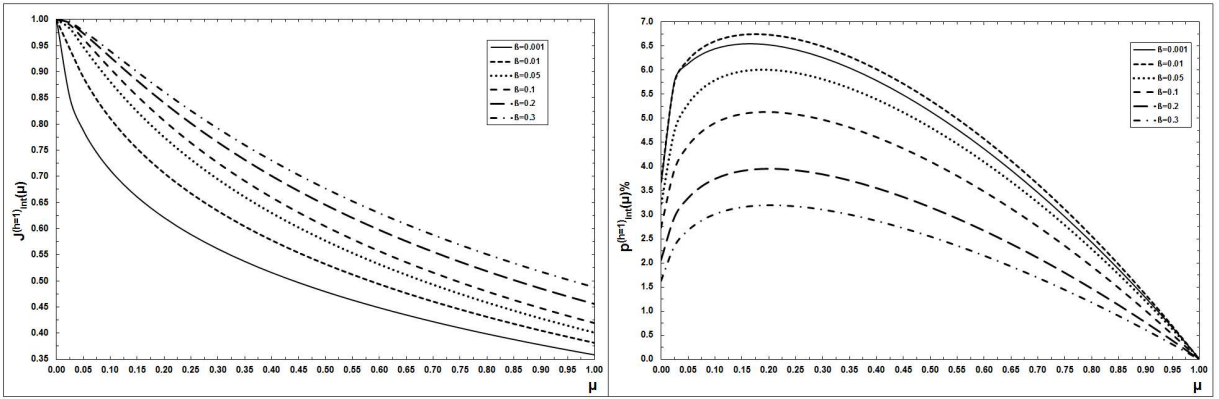}
\caption{Angular distribution $J_{int}(\mu)$ and degree of polarization $p_{int}(\mu)\%$ for exponential source $\exp{(-\tau)}$  for different values of parameter $\beta$. Positive $p_{int}(\mu)$ corresponds to wave electric field oscillations  parallel to the plane $({\bf nN})$ - the line of sight ${\bf n}$ and the normal to accretion disc ${\bf N}$ . }
 \label{a}
\end{figure*}

\begin{figure*}[t!]
\includegraphics[width=16cm]{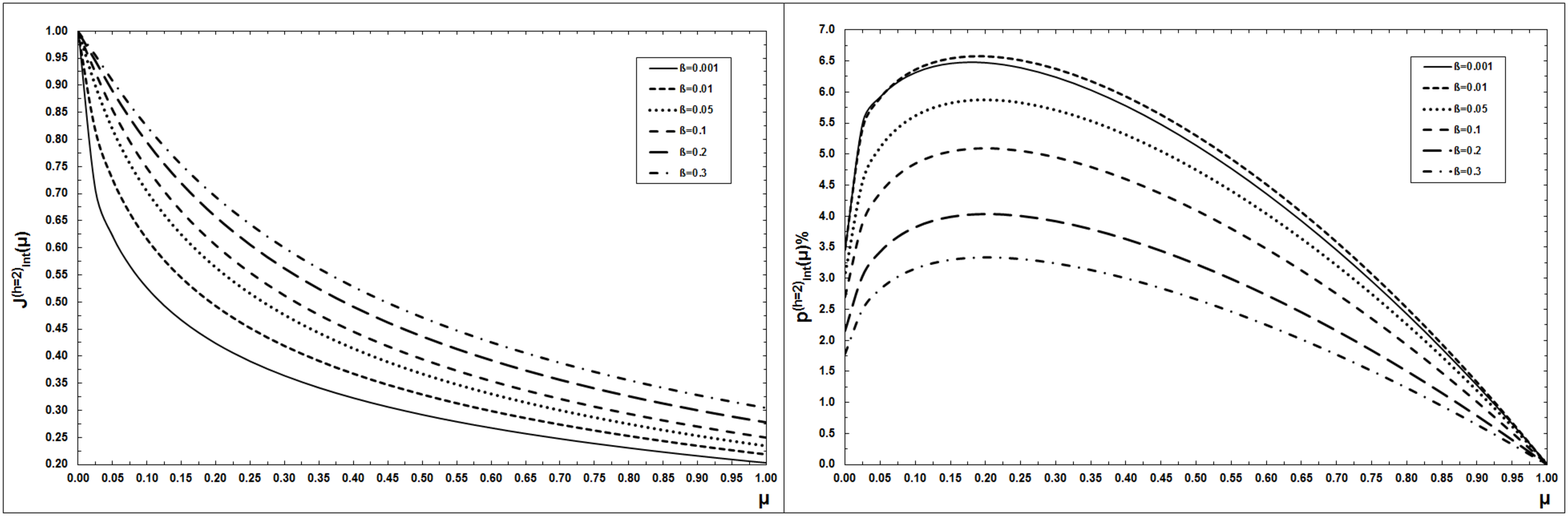}
\caption{Angular distribution $J_{int}(\mu)$ and degree of polarization $p_{int}(\mu)\%$ for exponential source $\exp{(-2\tau)}$  for different values of parameter $\beta$. Positive $p_{int}(\mu)$ corresponds to wave electric field oscillations  parallel to the plane $({\bf nN})$ - the line of sight ${\bf n}$ and the normal to accretion disc ${\bf N}$ . }
 \label{a}
\end{figure*}

For exponential sources $\sim \exp{(-\tau)}$ and

 $\sim \exp{(-2\tau)}$ the density of radiation $S_I(\tau)$ is located near the surface
of the medium. That is why the angular distributions $J_{int}(\mu)$ have the maximum values at $\mu=0$. The incident radiation falls on the  resonant atoms mostly perpendicular to the normal ${\bf N}$.  As a result, the polarization of emerging radiation is
positive, i.e. the wave electric field oscillates in the plane $({\bf nN})$. The degrees of polarization $p_{int}(\mu)$ have the
maximum values at $\mu\simeq 0.15\div0.2$ ($\vartheta\simeq 80.4^{\circ}\div 78.4^{\circ}$). The angular dependence and
polarization for sources $\exp{(-\tau)}$ and $\exp{(-2\tau)}$ are close  with one another. This is due to diffusion of the radiadion which
gives rise to the  smoothing  of the radiation  density from initial sources located near the surface of the medium.

It is of interest that for homogeneous source at $\beta=0$ the value ${\bf I}(x,0)$ is independent of the frequency $x$.
This follows from Eq.(37), if we recall that $\hat H_0(0)=\hat A(0)=\hat E$ and  $\alpha(x)=\varphi(x)+\beta$. 

 Tables 4 and 5 present
 $J(x,\mu),\,\, p(x,\mu)$, for homogeneous, linearly increased and exponential  (h=1) sources  
at $\mu=0.5\, (\vartheta=60^{\circ})$ for $\beta=0.01$  and  $\beta=0.1$, correspondingly. These tables can be useful for  estimates of polarization of spectral line radiation emitted from optically thick accretion discs.

 For the same aim we give the Tables 6 and 7, which characterize the intensity and polarization in the centre
of resonance line $x=0$. In these Tables the value  $J(\mu)=I(0,\mu)/I(0,0)$ characterizes the angular distribution of
emerging radiation  and $p(\mu)=Q(0,\mu)/I(0,\mu)$ is the polarization degree of this radiation.

\begin{table}[t]
\caption{ The values of $I_{int}(0)$  for a number of considered types of sources .}
\small
\begin{tabular}{l l l l l}
\hline
$\beta$       & $I^{(0)}_{int}(0)$     & $I^{(1)}_{int}(0)$   & $I^{(h=1)}_{int}(0)$  & $I^{(h=2)}_{int}(0)$\\
\hline
  0.001   & 34.300   & 2268.    & 4.059     & 3.416     \\
  0.01    & 9.720    & 90.99       & 3.055    & 2.619     \\
  0.05    & 3.786    & 8.588       & 2.112    & 1.874    \\
  0.1      & 2.438    & 2.812       & 1.656    & 1.503   \\
  0.2      & 1.518    & 0.822       & 1.206    & 1.122     \\
  0.3      & 1.129    & 0.374       & 0.962    & 0.908    \\
  0.4      & 0.906    & 0.206       & 0.804    & 0.766   \\
  0.5      & 0.750    & 0.127       & 0.691    & 0.664    \\
  0.6      & 0.655    & 0.084       & 0.607    & 0.587    \\  
  0.7      & 0.577    & 0.059       & 0.542    & 0.525   \\
  0.8      & 0.515    & 0.043       & 0.489    & 0.476     \\
  0.9      & 0.466    & 0.033       & 0.445    & 0.435    \\
\hline
\end{tabular}
\end{table}

\begin{table}[h!]
\caption{ The values of $J^{(0)}(x,\mu)$, $p^{(0)}(x,\mu)\%$, $J^{(1)}(x,\mu)$, $-p^{(1)}(x,\mu)\%$,  $J^{(h=1)}(x,\mu)$,
$p^{(h=1)}(x,\mu)\% $ at $\beta=0.01$ and  $\mu=0.5$  }
\small
\begin{tabular}{p{0.5cm}|p{0.6cm}p{1.0cm}|p{0.6cm}p{0.9cm}|p{0.6cm}p{0.9cm}}
\hline
$x$ & $J^{(0)}$ & $p^{(0)}$ & $J^{(1)}$ & $-p^{(1)}$ & $J^{(h=1)}$ & $p^{(h=1)}$ \\
\hline
0.  & 1.        & 0.694 & 1.    & 3.816     & 1.   & 5.737    \\
0.1 & 1.002 & 0.704 & 1.003 & 3.798  & 0.998  & 5.740   \\
0.2 & 1.009 & 0.731 & 1.012 & 3.741  & 0.990  & 5.748   \\
0.3 & 1.020 & 0.775 & 1.028 & 3.648  & 0.978  & 5.759   \\
0.4 & 1.037 & 0.834 & 1.052 & 3.517  & 0.961  & 5.770    \\
0.5 & 1.059 & 0.905 & 1.084 & 3.352  & 0.937  & 5.777    \\
0.6 & 1.087 & 0.983 & 1.126 & 3.153  & 0.908  & 5.774   \\
0.7 & 1.122 & 1.063 & 1.180 & 2.924  & 0.872  & 5.755    \\
0.8 & 1.165 & 1.139 & 1.250 & 2.669  & 0.829  & 5.713    \\
0.9 & 1.216 & 1.203 & 1.339 & 2.395  & 0.779  & 5.637   \\
1.0 & 1.276 & 1.246 & 1.453 & 2.108  & 0.723  & 5.522   \\
1.1 & 1.345 & 1.264 & 1.597 & 1.817  & 0.660  & 5.362   \\
1.2 & 1.421 & 1.250 & 1.781 & 1.531  & 0.592  & 5.156   \\
1.3 & 1.501 & 1.204 & 2.012 & 1.259  & 0.520  & 4.907   \\
1.4 & 1.580 & 1.129 & 2.302 & 1.010  & 0.447  & 4.625   \\
1.5 & 1.648 & 1.030 & 2.657 & 0.789  & 0.374  & 4.324    \\
1.6 & 1.691 & 0.918 & 3.073 & 0.602  & 0.305  & 4.020   \\
1.7 & 1.694 & 0.801 & 3.529 & 0.450  & 0.241  & 3.731   \\
1.8 & 1.640 & 0.690 & 3.969 & 0.332  & 0.184  & 3.471   \\
1.9 & 1.521 & 0.591 & 4.298 & 0.244  & 0.136  & 3.249  \\
2.0 & 1.339 & 0.509 & 4.400 & 0.183  & 0.097  & 3.069   \\
2.1 & 1.110 & 0.444 & 4.188 & 0.141  & 0.067  & 2.930   \\
2.2 & 0.864 & 0.395 & 3.668 & 0.113  & 0.045  & 2.827    \\
2.3 & 0.632 & 0.361 & 2.947 & 0.095  & 0.029  & 2.754   \\
2.4 & 0.436 & 0.337 & 2.182 & 0.083  & 0.019  & 2.705   \\
2.5 & 0.286 & 0.321 & 1.504 & 0.076  & 0.011  & 2.672   \\
2.6 & 0.180 & 0.310 & 0.977 & 0.072  & 0.007  & 2.651  \\
2.7 & 0.109 & 0.304 & 0.605 & 0.069  & 0.004  & 2.638  \\
2.8 & 0.064 & 0.300 & 0.360 & 0.067  & 0.002  & 2.630  \\
2.9 & 0.037 & 0.298 & 0.207 & 0.066  & 0.001  & 2.625  \\
3.0 & 0.020 & 0.296 & 0.116 & 0.066  & 0.001  & 2.622   \\
\hline
\end{tabular}
\end{table}

\begin{table}[h!]
\caption{Is similar to  Table 4 for $\beta=0.1$ .}
\small
\begin{tabular}{p{0.5cm}|p{0.6cm}p{1.0cm}|p{0.6cm}p{0.9cm}|p{0.6cm}p{0.9cm}}
\hline
$x$ & $J^{(0)}$ & $p^{(0)}$ & $J^{(1)}$ & $-p^{(1)}$ & $J^{(h=1)}$ & $p^{(h=1)}$ \\
\hline
0.  & 1.        & 1.682  & 1.    & 4.329     & 1.     & 4.230    \\
0.1 & 1.       & 1.684  & 1.003 & 4.304  & 0.996  & 4.230   \\
0.2 & 0.999 & 1.690  & 1.012 & 4.229  & 0.984  & 4.232   \\
0.3 & 0.997 & 1.699  & 1.027 & 4.105  & 0.964  & 4.234    \\
0.4 & 0.994 & 1.709  & 1.049 & 3.934  & 0.936  & 4.234    \\
0.5 & 0.989 & 1.719  & 1.077 & 3.720  & 0.899  & 4.231    \\
0.6 & 0.981 & 1.725  & 1.112 & 3.469  & 0.854  & 4.221    \\
0.7 & 0.969 & 1.725  & 1.153 & 3.186  & 0.800  & 4.203    \\
0.8 & 0.951 & 1.716  & 1.199 & 2.882  & 0.738  & 4.173   \\
0.9 & 0.927 & 1.696  & 1.247 & 2.566  & 0.670  & 4.129   \\
1.0 & 0.893 & 1.664  & 1.294 & 2.252  & 0.596  & 4.070  \\
1.1 & 0.847 & 1.619  & 1.332 & 1.951  & 0.518  & 3.998   \\
1.2 & 0.789 & 1.563  & 1.352 & 1.675  & 0.440  & 3.915  \\
1.3 & 0.718 & 1.499  & 1.345 & 1.430  & 0.365  & 3.825   \\
1.4 & 0.635 & 1.431  & 1.300 & 1.221  & 0.294  & 3.734   \\
1.5 & 0.544 &  1.365 & 1.211 & 1.050  & 0.230  & 3.645    \\
1.6 & 0.449 &  1.303 & 1.080 & 0.914  & 0.176  & 3.565  \\
1.7 & 0.357 &  1.250 & 0.919 & 0.809  & 0.130  & 3.495  \\
1.8 & 0.273 &  1.205 & 0.744 & 0.731  & 0.094  & 3.438   \\
1.9 & 0.201 &  1.169 & 0.573 & 0.673  & 0.066  & 3.393  \\
2.0 & 0.143 &  1.142 & 0.422 & 0.632  & 0.045  & 3.358   \\
2.1 & 0.099 &  1.122 & 0.298 & 0.603  & 0.030  & 3.333  \\
2.2 & 0.066 &  1.108 & 0.202 & 0.584  & 0.020  & 3.316    \\
2.3 & 0.043 &  1.098 & 0.133 & 0.571  & 0.013  & 3.303   \\
2.4 & 0.027 &  1.092 & 0.085 & 0.562  & 0.008  & 3.295   \\
2.5 & 0.017 &  1.088 & 0.053 & 0.556  & 0.005  & 3.290   \\
2.6 & 0.010 &  1.085 & 0.032 & 0.553  & 0.003  & 3.287   \\
2.7 & 0.006 &  1.085 & 0.019 & 0.550  & 0.002  & 3.285  \\
2.8 & 0.003 &  1.083 & 0.011 & 0.549  & 0.001  & 3.284   \\
2.9 & 0.002 &  1.082 & 0.006 & 0.548  & 0.001  & 3.283   \\
3.0 & 0.001 &  1.081 & 0.003 & 0.548  & 0.000  & 3.282   \\
\hline
\end{tabular}
\end{table}

\begin{table}[h!]
\caption{ The values $J^{(0)}(\mu)$   $p^{(0)}(\mu)\%$,
 $J^{(1)}(\mu)$, $-p^{(1)}(\mu)\%$  and $J^{(h=1)}(\mu)$,
$p^{(h=1)}(\mu)\% $ for $\beta=0.01$ and  $x=0$ (the centre of line).}
\small
\begin{tabular}{p{0.5cm}|p{0.6cm}p{1.0cm}|p{0.6cm}p{0.9cm}|p{0.6cm}p{0.9cm}}
\hline
$\mu$ & $J^{(0)}$ & $p^{(0)}$ & $J^{(1)}$ & $-p^{(1)}$ & $J^{(h=1)}$ & $p^{(h=1)}$\\
\hline
0.     & 1.       & -4.163 & 1.       & 12.108 & 1.        & 3.690    \\
0.05 & 1.089 & -2.430 & 1.099 & 10.116 & 1.002  & 5.231   \\
0.1   & 1.155 & -1.530 & 1.176 & 8.924  & 0.983  & 5.915    \\
0.15 & 1.192 & -0.904 & 1.247 & 7.986  & 0.960  & 6.307    \\
0.2   & 1.265 & -0.441 & 1.315 & 7.195  & 0.936  & 6.517    \\
0.25 & 1.313 & -0.090 & 1.380 & 6.502  & 0.911  & 6.597    \\
0.3   & 1.359 &  0.176 & 1.444 & 5.878  & 0.887  & 6.575    \\
0.35 & 1.402 &  0.378 & 1.506 & 5.308  & 0.863  & 6.468    \\
0.4   & 1.443 &  0.525 & 1.567 & 4.528  & 0.840  & 6.287   \\
0.45 & 1.481 &  0.629 & 1.627 & 4.047  & 0.818  & 6.042   \\
0.5   & 1.518 &  0.694 & 1.687 & 3.816  & 0.792  & 5.737   \\
0.55 & 1.553 &  0.727 & 1.746 & 3.371  & 0.776  & 5.379   \\
0.6   & 1.587 &  0.731 & 1.804 & 2.946  & 0.757  & 4.968  \\
0.65 & 1.619 &  0.708 & 1.862 & 2.537  & 0.737  & 4.508  \\
0.7   & 1.650 &  0.663 & 1.920 & 2.142  & 0.719  & 4.001   \\
0.75 & 1.680 &  0.597 & 1.978 & 1.760  & 0.701  & 3.448    \\
0.8   & 1.708 &  0.511 & 2.036 & 1.390  & 0.684  & 2.849  \\
0.85 & 1.736 &  0.407 & 2.093 & 1.028  & 0.666  & 2.205  \\
0.9   & 1.762 &  0.287 & 2.150 & 0.678  & 0.650  & 1.151   \\
0.95 & 1.787 &  0.151 & 2.207 & 0.335  & 0.634  & 0.780  \\
1.     & 1.811 &  0.       & 2.264 & 0.        & 0.619  & 0.         \\
\hline
\end{tabular}
\end{table}

\begin{table}[h!]
\caption{ Is similar to Table 6 for $\beta=0.1$. }
\small
\begin{tabular}{p{0.5cm}|p{0.6cm}p{1.0cm}|p{0.6cm}p{0.9cm}|p{0.6cm}p{0.9cm}}
\hline
$\mu$ & $I^{(0)}$ & $p^{(0)}$ & $I^{(1)}$ & $-p^{(1)}$ & $I^{(h=1)}$ & $p^{(h=1)}$\\
\hline
0.     & 1.       &-0.935  & 1.      & 15.664 & 1.      & 2.735    \\
0.05 & 1.067 & 0.365 & 1.136 & 13.254 & 0.993 & 3.977   \\
0.1   & 1.113 & 0.958 & 1.258 & 11.611 & 0.967 & 4.499    \\
0.15 & 1.151 & 1.325 & 1.378 & 10.262 & 0.938 & 4.784    \\
0.2   & 1.184 & 1.559 & 1.498 & 9.104  & 0.909  & 4.925    \\
0.25 & 1.213 & 1.703 & 1.618 & 8.089  & 0.879  & 4.964   \\
0.3   & 1.239 & 1.782 & 1.738 & 7.184  & 0.851  & 4.927   \\
0.35 & 1.263 & 1.811 & 1.860 & 6.369  & 0.823  & 4.826   \\
0.4   & 1.284 & 1.800 & 1.982 & 5.628  & 0.797  & 4.672   \\
0.45 & 1.304 & 1.755 & 2.106 & 4.951  & 0.772  & 4.472   \\
0.5   & 1.332 & 1.682 & 2.230 & 4.329  & 0.748  & 4.230   \\
0.55 & 1.339 & 1.585 & 2.355 & 3.754  & 0.725  & 3.950  \\
0.6   & 1.354 & 1.468 & 2.482 & 3.220  & 0.703  & 3.634   \\
0.65 & 1.369 & 1.332 & 2.608 & 2.723  & 0.682  & 3.286   \\
0.7   & 1.382 & 1.180 & 2.736 & 2.259  & 0.662  & 2.906   \\
0.75 & 1.394 & 1.013 & 2.865 & 1.824  & 0.642  & 2.494    \\
0.8   & 1.406 & 0.833 & 2.994 & 1.416  & 0.624  & 2.054   \\
0.85 & 1.416 & 0.641 & 3.125 & 1.031  & 0.606  & 1.583   \\
0.9   & 1.426 & 0.438 & 3.260 & 0.668  & 0.589  & 1.084   \\
0.95 & 1.435 & 0.224 & 3.387 & 0.325  & 0.573  & 0.556  \\
1.     & 1.443 & 0.       & 3.519 & 0.        & 0.557 & 0.        \\
\hline
\end{tabular}
\end{table}

\section{Conclusion}

  In this paper we solve the  radiative transfer equation for polarized  resonance line in three component medium, consisting of
resonant atoms, non-resonant atoms and grain particles. This equation (see Eq.(15)) generalizes usual transfer equation for medium consisting of resonant atoms only. The first term of this equation has the term $\alpha(\nu)=\varphi(\nu)+\beta$, where $\varphi(\nu)$ describes the normalized shape of scattering and absorption cross-sections, and parameter $\beta$ describes the line intensity extinction due to existence of grains and non-resonant atoms. The factor $(1-\epsilon)$ in second (integral) term is equal to probability of scattering   t on resonant atom. Usually $\epsilon\sim 10^{-4}$ is much smaller than parameter $\beta$. For this reason we take $\epsilon=0$ and study the dependence of radiation intensity and polarization on parameter $\beta$.

    We derived the solution of transfer equation using
the matrix resolvent technique. This method allows us to obtain the intensity and polarization of radiation outgoing  from semi-infinite
atmosphere for sources of non-polarized radiation of homogeneous, linearly increasing and exponentially decreasing  types. To
obtain these values one must to solve the system of non-linear equations like those for Chandrasekhar H-functions. 

\section{\bf Acknowledgements.}
This research was supported by the Basic Research Program P-7 of Prezidium of Russian Academy of Sciences and 
the Program of the Department of Physical Sciences of Russian
Academy of Sciences No 2.

 We are very grateful to anonymous referee for numerous useful remarks.


\begin{thebibliography}{60}
\bibitem{1} Chandrasekhar, S.: Radiative transfer. Dover, New York (1960)
\bibitem{2} Dementyev, A. V. Astron. L. {\bf 34}, 574 (2008)
\bibitem{3} Dementyev, A. V. Ap {\bf 53}, 419 (2010)
\bibitem{4} Dolginov, A. Z., Gnedin, Yu. N., Silant'ev, N. A. : Propagation and Polarization of
                    Radiation in Cosmic Media. Gordon \&  Breach Publ., Amsterdam (1995)
\bibitem{5} Faurobert, M. A\& A {\bf 194}, 268 (1988)
\bibitem{6} Faurobert-Sholl, M. , Frisch H. A\& A {\bf 219}, 338 (1989)
\bibitem{7} Faurobert-Sholl, M., Frisch, H., Nagendra, K. N. A\& A {\bf 322}, 896 (1997)
\bibitem{8} Fluri,D. M.: Radiative transfer with polarized
 scattering in the magnetized Solar atmosphere. Cuvillier Verlag, Gottingen (2003)
\bibitem{9} Frisch, U., Frisch, H., MNRAS, {\bf 181}, 273 (1977)
\bibitem{10} Gasiorovich, S.: Quantum physics. John Wiley \& Sons, NY (1996)
\bibitem{11} Hummer, D. G., Rybicki, G., AR A\&A, {\bf 9}, 237 (1971)
\bibitem{12} Ivanov V.V. Radiative transfer in spectral lines. National Bureau of Standarts, Washington, (1973) (translation from russian    
 edition 1969)
\bibitem{13} Ivanov, V.V.,  A\&A {\bf 307}, 319 (1996)
\bibitem{14} Ivanov, V.V., Grachev S.I., Loskutov, V. M.  A\&A {\bf 318}, 609 (1997)
\bibitem{15} Kriese, J. T., Siewert, C. E. ,  Astrophys. J. {\bf 164}, 389 (1971)
\bibitem{16} Landau \& Lifshitz, Course of Theorethical Physics, v.4, 1968 (translated as Berestetskii V.B., Lifshitz E. M.,Pitaevskii L. P.,   
  Relativistic quantum theory.Pergamon Press, Oxford (1971)
\bibitem{17} Landi degl'Innocenti E., Landolfi, M. : Polarization in Spectral Lines. Dordrecht: Kluwer Acad. Publ. (2004)
\bibitem{18} McKenna, S.J. ApSS, {\bf 108}, 31 (1985)
\bibitem{19} Mihalas D.: Stellar atmospheres. W.H.Freeman and Company, San Francisco (1982)
\bibitem{20} Rees, D. E., Saliba, G., A\&A, {\bf 115}, 1, (1982)
\bibitem{21} Rooij, de W. A., Bosma, P. B., Hooff, van , J.P.C.,  A\&A {\bf  226}, 347  (1989)
\bibitem{22} Silant'ev, N. A., Astron. Report, {\bf 57 }, 587 (1980)
\bibitem{23} Silant'ev, N. A., Alekseeva, G. A., Novikov, V. V. Ap\&SS {\bf 357}, 53 (2015)
\bibitem{24} Smirnov, V.I.: The course of higher mathematics, Vol. 4. Integral equations and Partial differential equations, Pergamon Press,     
 New York (1964)
\bibitem{25} Sobolev, V. V.: Course in theoretical astrophysics. NASA Technical Translation F-531, Washington (1969)
\bibitem{26} Stenflo, J.O. A\&A, {\bf 46}, 61 (1976)
\end{thebibliography}
\end{document}